\documentclass[aps,twocolumn,floats,nofootinbib]{revtex4} 
\usepackage{amsmath}
\usepackage{graphics,graphicx,float}
\usepackage{amssymb,color,bm,dcolumn,accents}
\usepackage{epsf,epstopdf,wrapfig}
\usepackage{textgreek}
\usepackage{ulem}

\newcommand{\beq}{\begin{equation}}
\newcommand{\eeq}{\end{equation}}
\newcommand{\beqs}{\begin{equation*}}
\newcommand{\eeqs}{\end{equation*}}
\newcommand{\beqn}{\begin{eqnarray}}
\newcommand{\eeqn}{\end{eqnarray}}
\newcommand{\beqns}{\begin{eqnarray*}}
\newcommand{\eeqns}{\end{eqnarray*}}

\newcommand{\Tr}{\text{Tr}}

\newcommand{\RV}{\color[rgb]{0,0,0}}

\begin{document}

\title{Searching for long time scales without fine tuning}
\author{Xiaowen Chen$^{1}$\footnote{xiaowenc@princeton.edu}\footnote{Present address: Laboratoire de Physique de l'Ecole Normale Sup\'erieure, ENS, PSL Universit\'e, CNRS,  Sorbonne Universit\'e, Universit\'e Paris Cit\'e, F-75005 Paris, France. \\ {xiaowen.chen@phys.ens.fr}} and William Bialek$^{1,2}$\footnote{wbialek@princeton.edu}}

\affiliation{$^1$Joseph Henry Laboratories of Physics, and Lewis--Sigler Institute for Integrative Genomics, Princeton University, Princeton NJ 08544\\
$^2$Initiative for the Theoretical Sciences, The Graduate Center, City University of New York, 365 Fifth Ave, New York NY 10016}
\date{\today}

\begin{abstract}
Animal behavior occurs on time scales much longer than the response times of individual neurons.  In many cases, it is plausible that these long time scales emerge from the recurrent dynamics of electrical activity in networks of neurons.  In linear models, time scales are set by the eigenvalues of a dynamical matrix whose elements measure the strengths of synaptic connections between neurons.  It is not clear to what extent these matrix elements need to be tuned in order to generate  long time scales; in some cases, one needs not just a single long time scale but a whole range.  Starting from the simplest case of random symmetric connections, we combine maximum entropy and random matrix theory methods to construct ensembles of networks, exploring the constraints required for long time scales to become generic.  We argue that a single long time scale can emerge generically from realistic constraints, but a full spectrum of slow modes requires more tuning. Langevin dynamics that generates patterns of synaptic connections drawn from these ensembles involves a combination of Hebbian learning and activity--dependent synaptic scaling.
\end{abstract}

\maketitle

\section{Introduction}
\label{sec:lts_intro:intro}

Living systems face various challenges over their lifetimes, and responding to these challenges often involves  behaviors that occur over multiple time scales~\cite{bialek2024long}.  
As an example, a migratory bird needs both to react to instantaneous gusts and to navigate its course over months~\cite{flack2018local}. Thanks to recent development of behavior tracking and image analysis techniques, long time scales have been identified and quantified in the behaviors of many animals, including the scale invariant correlations spanning three decades in time in the fruit fly~\cite{bialek2024long}, the near--marginal modes in locally linear approximations to neural and behavioral dynamics of the nematode \textit{C. elegans}~\cite{costa_adaptive_2019}, and the non-Markovian behavior in the larval zebrafish~\cite{sridhar2024uncovering}.

How do those time scales emerge from the generator of the behavior, the nervous system? While transient responses of individual neurons decay on the time scale of tens of milliseconds, autonomous behaviors can last orders of magnitude longer.  We see this when we hold a string of  numbers in our heads for tens of seconds before dialing a phone, and when a musician plays a piece from memory that lasts many minutes. 
While this emergence of long time scales can arise from the coupling of living systems to a slowly fluctuating environment~\cite{costa2024fluctuating, morrell2021latent, schwab2014zipf} such as the day-night cycle, experimental evidences support that long time scales can be generated by interactions of the neurons. 
On one hand, single long time scales in behavior have been associated with persistent neural activities, where after a pulse stimulation, some neurons are found to hold their firing rate at specific values that encode the transient stimuli~\cite{aksay_vivo_2001, brody_timing_2003, major_plasticity_2004, major_plasticity_2004-1, major_persistent_2004, srimal_persistent_2008}.
On the other hand, a wide range of persistent time scales has been observed, e.g. in various cortical areas of monkeys~\cite{murray2014hierarchy, bernacchia2011reservoir} and macaque prefrontal cortex~\cite{rigotti2010internal}, as well as within the same local circuit such as the larval zebrafish oculomotor system~\cite{miri2011spatial}. Moreover, in the mouse hippocampus, coarse-graining neuronal activities has uncovered reproducible scaling behaviors of networks of 1000+ neurons~\cite{meshulam2019coarse}. As verified both biologically and computationally, these multidimensional dynamics allow the network to integrate information at a broad range of timescales, which are beneficial for decision making~\cite{chaudhuri2015large}, memory storage~\cite{fusi2005cascade}, task switching~\cite{rigotti2010internal}, and better prediction of multiscale dynamics~\cite{tanaka2022reservoir}.  All of these observations motivate a search for mechanisms allowing the emergence of long time scales.

It is plausible that long time scales emerge from the recurrent dynamics of electrical activity in the network of neurons.
 In the simplest linear model, the relaxation times of the system depend on the eigenvalues of a matrix representing the synaptic connection strengths among neurons, and we can imagine this being tuned so that time scales become arbitrarily long~\cite{seung_how_1996}.  This simple model has  successfully explained the long time scale in the oculomotor system of goldfish, where the nervous system tunes its dynamics to be slow and stable using constant feedback from the environments~\cite{major_plasticity_2004, major_plasticity_2004-1}. 
In general, long time scales in linear dynamical systems require fine tuning, as the modes need to be slow, but not unstable.  

To circumvent this biologically-implausible fine tuning requirement, current approaches of modeling interacting networks which generate long time scales include either changing the setup completely by adding in non-linearity~\cite{brody_basic_2003}, fine tuning the structure of the networks using biological structures~\cite{chaudhuri2015large} or highly structured networks such as feedforward~\cite{goldman_memory_2009}, ring~\cite{burak_fundamental_2012}, or grid networks~\cite{shi2023spatial}, imposing clustered connections~\cite{litwin2012slow}, adding hetereogeneity to the self-interaction~\cite{stern2023reservoir}, or promoting the interaction matrix to dynamical variables regulating the overall neural activity~\cite{magnasco_self-tuned_2009, renart_robust_2003, tetzlaff_synaptic_2013}.
Nonetheless, it remains unclear how much fine-tuning of the interactions is required to generate long time scales. If we allow all possible connections, what constraints on the strengths of these connections are required for the emergence of long time scales?  We can ask this question both for the case of a single long time scale and for a  continuum spectrum time scales, as observed in animal behaviors~\cite{bill_dimension_dynamics}.

In this manuscript, we address the fine-tuning question by asking whether we can find ensembles of random connection matrices, subject to biologically plausible constraints, such that the resulting time scales of the system grow with increasing system sizes. We also discuss the conditions for systems to exhibit a continuous spectrum of slow modes as opposed to single slow modes. Finally, we present plausible dynamics for the system to tune its connection matrix towards these desired ensembles, with the hope that these dynamics may correspond to actual regulating process in real biological systems.

\section{Setup}
\label{sec:lts_intro:rmt}

The problem of characterizing time scales in fully nonlinear neural networks (or any high dimensional dynamical system) is very challenging.  To make progress, we follow the example of Ref.~\cite{seung_how_1996} and consider the case of linear networks, as to approximate the dynamics near fixed points.  For linear systems, time scales are related to the eigenvalues of the dynamical matrix that embodies the pattern of synaptic connectivity.  The question of whether behavior is generic can be made precise by drawing these  matrices at random from some probability distribution, connecting with the large literature on random matrix theory \cite{livan_introduction_2018,marino_number_2016}.  Importantly, we expect that some behaviors in these ensembles of networks become sharp as the networks become large, a result which has been exploited in thinking about problems ranging from energy levels of quantum systems~\cite{wigner_on_1951} to ecology~\cite{may_will_1972} and finance~\cite{bouchaud_theory_2003}.

Concretely, we represent the activity each neuron $i = 1, 2, \cdots , N$ by a continuous variable $x_i$, which we might think of as a smoothed version of the sequence of action potentials, and assume a linear dynamics
\beq\label{eq:lin_dyn_wn}
\dot{x_i} = - x_i + \sum_j M_{ij} x_j + \eta_i(t)\,\mbox{.}
\eeq
{See Fig.~\ref{fig:setup}(a) for schematics. For simplicity, we assume the interaction matrix (i.e. the synaptic strength) $\bm{M}$ is symmetric, i.e. $M_{ij} = M_{ji}$ (see Sec.~\ref{sec:sym_vs_general_m} and Appendix E for discussions on general $\bm M$ without symmetry constraint).  
If the neurons were unconnected ($\bm{M}=0$), their activity $x$ would relax exponentially on a time scale which we choose as our unit of time.  In what follows it will be important to imagine that the system is driven, at least weakly, and we take these driving terms to be independent in each cell and uncorrelated in time,
where $\langle \eta_i(t) \rangle = 0$ and
\begin{equation}
\langle \eta_i(t)\eta_j(t') \rangle = 2\delta_{ij}\delta(t-t').
\label{noise1}
\end{equation}
The choice of white noise is conventional, but also important because we want to understand how time scales emerge from the network dynamics rather than being imposed upon the network by outside inputs.

In linear systems we can rotate to independent modes, corresponding to weighted combinations of the original variables.  If the matrix $\bm{M}$ is symmetric, then the dynamics are described by the relaxation times of these modes, 
\beq
\tau_i \equiv \frac{1}{1-\lambda_i} = \frac{1}{k_i},
\eeq
where $\{\lambda_i\}$ are the eigenvalues of $\bm{M}$; the system is stable only if all $\lambda_i < 1$.

If the matrix $\bm{M}$ is chosen from a distribution $P(\bm{M})$ then the eigenvalues are random variables, but their density, for example, becomes smooth and well defined in the limit $N\rightarrow\infty$,
\begin{equation}
\rho(\lambda) \equiv \lim_{N\rightarrow\infty}\frac{1}{N}\sum_{i=1}^N \delta(\lambda - \lambda_i) .
\end{equation}
The width of the support, i.e. where the density of eigenvalues is nonzero, is denoted by $l$. This density of eigenvalues, $\rho(\lambda)$, is the spectrum of the interaction matrix $\bm{M}$, and determines the timescales in the system (see Fig.~\ref{fig:setup}(b).) Systems with long time scales are defined such that the time scales grow as the system size $N$ {\RV{or faster}}. Specifically, we focus on two types of time scales: 

%%%%%%%%%%%%
\begin{figure}[ht]
\includegraphics[scale=1]{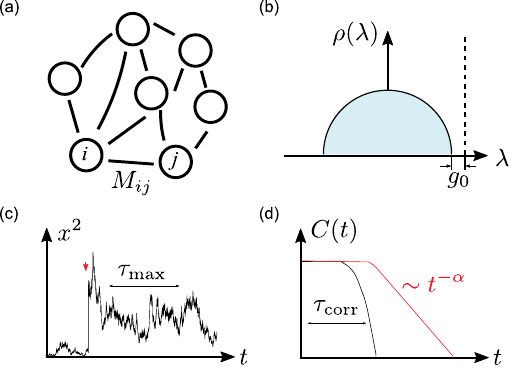}
\caption{Schematics for symmetric linear dynamical systems, and the corresponding timescales. (a) A linear dynamical system with damping and pairwise interaction $\bm{M}$
%. {\RV{(b) The eigenvalue spectrum of $\bm{M}$ determines the timescales of the dynamical system, especially the gap $g_0$ between the right edge of the support to the stability threshold. \sout{
has time scales determined by the eigenvalue spectrum of $\bm{M}$, especially the gap $g_0$ to the stability threshold (panel (b))%} }}. %
We focus on two timescales. 
(c) The \textit{longest time scale} is defined as $\tau_\text{max} = 1/g_0$, and gives the time scale for the norm activity to decay when the system is perturbed {\RV{at the time indicated by the \textit{red arrow}}}.
(d) The \textit{correlation time} $\tau_\text{corr}$ is the characteristic time scale of de-correlation when the system is unperturbed. 
In cases where the system has a continuous range of long time scales, the correlations decay as a power law (\textit{red curve}). 
}\label{fig:setup}
\end{figure}

%%%%%%%%%%%%

\textit{1. Longest time scale.} The longest time scale, $\tau_\text{max}$, is the time constant given by the slowest mode of the system, which dominates the dynamics after long enough times after a perturbation (Fig.~\ref{fig:setup}(c)). This time scale is determined by the gap, $g_0$, between the largest eigenvalue and the stability threshold, which with our choice of units is $w = 1$.  Mathematically, we define
\beq
\tau_\text{max} \equiv \frac{1}{g_0} = \frac{1}{1-\lambda_\text{max}} .
\eeq
In the thermodynamic limit, the gap is taken to be between the stability threshold and the right edge of the support of the spectral density\footnote{We note that this approximation is not ideal, as the spectral distribution of eigenvalues does not converge uniformly. In some cases, the fluctuation of the largest eigenvalue can be more meaningful than the average.} (see Fig.~\ref{fig:setup}(b)).  

\textit{2. Correlation time.} The correlation time scale is the time scale it takes for the system to de-correlate, i.e. the decaying time constant of the correlation function (Fig.~\ref{fig:setup}(d)). To get at the correlation time, let's take seriously the idea that the network is driven by noise.  Then $x(t)$ become a stochastic process, and from Eqs. (\ref{eq:lin_dyn_wn}) and (\ref{noise1}) we can calculate the correlation function
\beq
\begin{split}
C_N(t) & \equiv \frac{1}{N}\sum_i \langle x_i(0) x_i(t) \rangle \\
& = \frac{1}{N}\sum_{i}\frac{1}{1-\lambda_i} e^{-(1-\lambda_i) |t|} \\
& = \frac{1}{N}\sum \tau_i e^{-|t|/\tau_i}.
\label{CNt}
\end{split}
\eeq
The normalized correlation function
\beq
R_N(t) \equiv \frac{C_N(t)}{C_N(0)} = \frac{\sum_i \tau_i e^{-|t|/\tau_i}}{\sum_i \tau_i}
\eeq
has the intuitive behavior of starting at $R_N(0) = 1$ and decaying monotonically.  Then there is a natural definition of the correlation time, by analogy with single exponential decays,
\beq
\tau_\text{corr}(\lbrace\lambda\rbrace) \equiv \int_0^\infty dt\, R_N(t) = \frac{\sum_i \tau_i^2}{\sum_i \tau_i}.
\label{t_corr}
\eeq
In the thermodynamic limit, the autocorrelation coefficient $R(t)$ and the correlation time $\tau_\text{corr}$ becomes the ratio of two integrals over the eigenvalue density $\rho(\lambda )$.

Importantly, $\tau_\text{max}$ depends only on the largest eigenvalue, while $\tau_\text{corr}$ depends on the entire spectrum, and hence can be used to differentiate cases where the system is dominated by a single vs. a continuous spectrum of slow modes.  The two time scales satisfy $ \tau_\text{corr} \leq \tau_\text{max} $, with the equality assumed only when all eigenvalues are equal, i.e. the spectral density is a delta function at $\lambda = \lambda_\text{max}$.

With these definitions, we refine our goal as to find biologically plausible ensembles for the connection matrix $M$, such that the resulting stochastic linear dynamics has time scales, $\tau_\text{max}$ and $\tau_\text{corr}$, that are ``long,''   growing as a power of the system size $N$, perhaps even extensively. To avoid fine tuning, we will construct examples of such ensembles by imposing global constraints on measurable observables of the dynamical system. We then compute the spectral density and the corresponding time scales using a combination of mean-field theory and numerical sampling of finite systems.

\subsection{Symmetric vs. general choice of interaction matrix $\bm{M}$}
\label{sec:sym_vs_general_m}

The most general choice for the interaction matrix $\bm M$ is without the symmetry constraint. Nonetheless, there are a few problems with this generality. First, the eigenvalues can now be complex, and the dynamics is composed of different modes of damped oscillations, introducing extra measures of time scales. Second, the correlation function for general interaction matrix depends not only on the eigenvalues, but also on the eigenvectors, vastly complicating the calculation (see Appendix E). In fact, in later parts of this manuscript, when we introduce stability constraints to the interaction matrix, random matrices without symmetry constraints becomes analytically unsolvable, while numerical results using brute force Monte Carlo sampling are limited to small system size. As suggested by the numerical results, asymmetric random matrices will exhibit similar fundamentals as symmetric random matrices, such as where possible phase transitions occur, and whether one can obtain slow modes. However, the scaling coefficient is different (Appendix E). For starting somewhere, we propose to leave solving the dynamics for general interaction matrices for the future, and will focus our discussion in this manuscript on symmetric matrices.

\section{Time scales for ensembles with different global constraints}

To construct linear dynamical systems that generate long time scales without fine tuning individual parameters, we want to find probability distributions $P(\bm M)$ for the connection matrix such that time scales are long, on average.  We start with the Gaussian distribution, and gradually add constraints.  We will see that for the Gaussian distribution itself, there is a critical value for the scaled variance of the elements of $\bm{M}$, $c_\text{crit} = 1/\sqrt{2}$.   For $c < c_\text{crit}$ the system is stable but time scales are short, while for $c > c_\text{crit}$ the system is unstable.  Exactly at the critical point $c = c_\text{crit}$  time scales  are long in the sense that we have defined, diverging with system size.  The essential challenge is to find the weakest constraints on $P(\bm M)$ that make these long time scales generic.  Many of the results that we need along the way are known in the random matrix theory literature, but we will arrive at some new theoretical questions.

\begin{figure}
\includegraphics[scale=1]{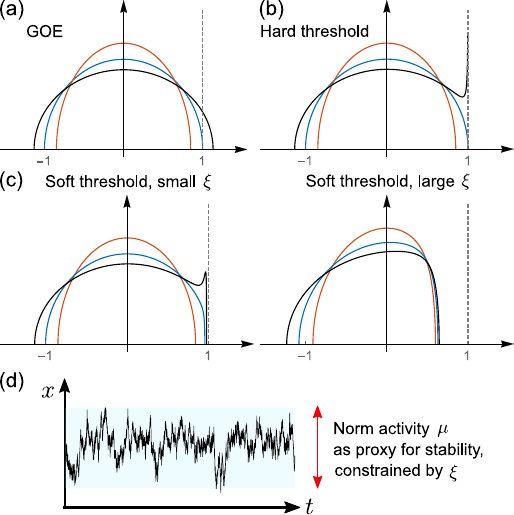}
\caption{ Spectral density for the connection matrix $\bm M$ drawn from {\RV{(a) Model 1,}} the Gaussian Orthogonal Ensemble (GOE); {\RV{(b) Model 2,}} the GOE with a hard threshold enforcing stability; and {\RV{(c) Model 3,}} the GOE with an additional global constraint on the norm activity.
Three representative parameters are chosen for each ensemble such that the system is subcritical ($c = 0.6$, \textit{red}), critical ($c = 1/\sqrt{2}$, \textit{blue}), and supercritical ($c = 0.8$, \textit{black}). The stability threshold is visualized as the dashed gray line at $\lambda_w = 1$. (d) Schematic for constraining the averaged norm activity {\RV{in Model 3}}, generating ensembles with the soft threshold.}\label{fig:spec}
\end{figure}

\subsection{Model 1: the Gaussian Orthogonal Ensemble}
\label{subsec:lts:goe_tscale}

The simplest ensemble for the interaction matrix $\bm M$ is the Gaussian Orthogonal Ensemble (GOE), which has been studied since the beginning of random matrix theory~\cite{wigner_on_1951, dyson_threefold_1962, dyson_brownianmotion_1962, livan_introduction_2018}. GOE has symmetric matrix elements, $M_{ij} = M_{ji}$, and the matrix elements are independent Gaussian random variables, with variances such that
\begin{eqnarray}
M_{ii}&\sim& \mathcal{N}(0,c^2/N),\\
M_{ij}\vert_{i\neq j} &\sim& \mathcal{N}(0, c^2/2N).
\end{eqnarray}
As the parameter $c$ sets the scale of connections, we will refer to this parameter as the interaction strength. The factor of $N$ in the variance ensures that the density $\rho(\lambda) $ has support over a range of eigenvalues that are  $\mathcal{O}(1)$ at large $N$. 
The eigenvalue distribution of the GOE was solved a long time ago~\cite{wigner_on_1951}. Here, we review these results, and use the eigenvalue distributions to characterize the longest and the correlation time scales. Despite its simplicity, Model 1 (GOE) provides a foundation for understanding the challenge of obtaining long time scales, before we introduce more complex random matrix ensembles.

The probability distribution for the interaction matrix $\bm M$ can be written as
\beq
P_\text{GOE}(\bm M)\propto \exp\left(
-\frac{N}{2c^2}\text{Tr} \bm M^\intercal \bm M
\right).
\label{GOE1}
\eeq 
Because the distribution only depends on matrix traces, it is invariant to rotations of $\bm M$. Equivalently, if we think of decomposing the matrix $\bm M$ into its eigenvectors and eigenvalues, the probability depends only on the eigenvalues. Thus, we can integrate out the eigenvectors, and obtain the joint distribution of eigenvalues (see Appendix A for details),
\beq
P_\text{GOE}(\lbrace\lambda_i \rbrace) \propto
\exp
\left[
-\frac{N}{2c^2}\sum_i\lambda_i^2
+\frac{1}{2}\sum_{j\neq k}\ln \big\lvert \lambda_j - \lambda_k \big \rvert
\right],
\eeq
where the logarithmic repulsion term emerges from the Jacobian when we change variables from matrix elements to eigenvalues and eigenvectors. For a bit more intuition, we can identify that this joint density function of eigenvalues is mathematically identical to the Boltzmann distribution of two-dimensional {\RV{Coulomb}} gas confined to one dimension {\RV{(the line of real numbers)}}, where each of the $N$ {\RV{eigenvalues}} experiences a{\RV{n external}} quadratic potential, and pairs of electrons repel logarithmically {\RV{(hence the two-dimensional Coulomb repulsion)}}.

The spectral density can then be found using a mean field approximation, which becomes exact  as $N\rightarrow\infty$. The result is Wigner's  well-known semicircle distribution~\cite{wigner_on_1951},
\beq
\rho_\text{GOE}(\lambda) = \frac{1}{\pi c}\sqrt{2 - \frac{\lambda^2}{c^2}}, \hspace{1cm} \lambda\in [-\sqrt{2}c, \sqrt{2}c].
\label{wigner1}
\eeq
The width of the support is $l = 2\sqrt{2}c$. For completeness, we review the derivation of the spectral density in Appendix A.

Equation (\ref{wigner1}) for the spectral density, together with Fig.~\ref{fig:spec}(a), shows that there is a phase transition at $c_\text{crit} = 1/\sqrt{2}$. If the interaction strength is greater than this critical strength (supercritical), then $\lambda_\text{max} > 1$ and the system becomes unstable. If the interaction strength is smaller (subcritical), then the gap size between the largest eigenvalue and the stability threshold $\lambda = 1$ is of order 1, and the time scales remain finite as system size increases. The only case when the system has slow modes is at the critical value of the interaction strength, $c = c_\text{crit} = 1/\sqrt{2}$, where the spectral density becomes tangential to the stability threshold. 

At criticality, corresponding to the blue curve in Fig.~\ref{fig:spec}(a), we can estimate the size of the gap  by asking that the gap be large enough to contain $\sim 1$ mode, that is
\begin{equation}
\int_{1-g_0}^1 d\lambda \,N \rho_\text{GOE}(\lambda) \sim 1 .
\label{gap1}
\end{equation}
Near $\lambda \rightarrow 1$, the spectral distribution can be approximated by $\rho_\text{GOE}(\lambda) \sim (1-\lambda)^{1/2}$, and this gives $Ng_0^{3/2} \sim 1$ or  $g_0 \sim N^{-2/3}$. Thus the longest time scale grows with system size, $\tau_\text{max}\sim N^{2/3}$. 

In the same way, we can estimate the full correlation function from the correlation time [Eq.~(\ref{CNt})],
\beq
C_N (t) = \frac{1}{N}\sum_i \tau_i e^{-t/\tau_i}\nonumber.
\eeq
In the thermodynamic limit, $N \rightarrow \infty$, 
\beq
\begin{split}
C_{N\rightarrow \infty}(t) & = \int d\lambda  {{\rho_\text{GOE}(\lambda) }\over{(1-\lambda)}} e^{-(1-\lambda) |t|} \\
 &\sim \int d\lambda  {{1 }\over{(1-\lambda)^{1/2}}} e^{-(1-\lambda) |t|} \\
& \sim |t|^{-1/2}. 
 \end{split}
\eeq
This has the power--law behavior expected for a critical system, where there is a continuum of slow modes.

Note that in the GOE system, slow modes with time scales growing as system size are only possible at a single value of the interaction strength. Nonetheless, we need to distinguish the fine tuning here as happening at an ensemble level, which is different from the element-wise fine tuning that might have been required if we considered particular interaction matrices. 

\subsection{Model 2: GOE with hard stability threshold}
\label{sec:lts_intro:hard}

Drawing interaction matrices from the GOE leads to long time scales only at a critical value of interaction strength. Can we modify the ensemble such that long time scales can be achieved without this fine tuning? In particular, in the GOE, if the interaction strength is too large, then the system becomes unstable. What will the spectral distribution look like if we allow $c> c_\text{crit}$ but edit out of the ensemble any matrix that leads to instability? 

Mathematically, a global constraint on the system stability requires all eigenvalues to be less than the stability threshold, $\lambda_w = 1$. This modifies the distribution of the {\RV{interaction matrices}} with a Heaviside step function:
\beq
P_\text{hard}(\bm M) \propto
P_\text{GOE}(\bm M) \prod_i \Theta(1-\lambda_i).
\eeq
Conceptually, what this model does is to pull matrices out of the GOE and discard them if they produce unstable dynamics; the distribution $P_\text{hard}(\lbrace\lambda \rbrace)$ describes the matrices that remain after this editing.  Importantly we do not introduce any extra structure, and in this sense $P_\text{hard}$ is a maximum entropy distribution, as discussed more fully below.

The spectral density $\rho_\text{hard}(\lambda)$ that follows from $P_\text{hard}$ was first found by Dean and Majumdar~\cite{dean_large_2006, dean_extreme_2008, majumdar_top_2014}.   Again, there is a phase transition depending on the interaction strength. For ensembles with interaction strength less than the critical value $c_\text{crit} = 1/\sqrt{2}$,  the stability threshold is away from the bulk spectrum, so the spectral density remains as Wigner's semicircle. On the other hand, if the interaction strength is greater than the critical value, the spectral density becomes
\beq\label{eq:hard_push_spectral}
\rho_\text{hard}(\lambda) = \frac{1}{c^2}\frac{\sqrt{\lambda + l^* - 1}}{2\pi \sqrt{1 - \lambda}} \left(
l^* - 2 \lambda
\right),
\eeq
where the width of the support
\beq
l^* = \frac{2}{3}\left(
1 + \sqrt{1 + 6 c^2}
\right).
\eeq

Building on these existing results for the spectral distribution, we now can ask for an interpretation of the emergent time scales.  As shown in Fig.~\ref{fig:spec}(b), the stability threshold acts as a wall pushing the eigenvalues to pile up. More precisely, near the stability threshold $\lambda = 1$ we have $\rho_\text{hard}(\lambda) \sim (1-\lambda)^{-1/2}$,  which [by the same argument as in Eq.~(\ref{gap1})] indicates that the longest time scale increases as system size with $\tau_\text{max}\sim N^2$.

The autocorrelation function also is dominated by the eigenvalues close to the stability threshold.  
Following steps detailed in Appendix B, we can write the resulting autocorrelation coefficient as
\beq
R_\text{hard}(t) = \frac{C(t)}{C(0)} 
= 1 - \sqrt{\pi} (\varepsilon t)^{1/2} + \varepsilon t  + \mathcal{O}((\varepsilon t)^2),
\eeq
{\RV{where $\varepsilon \sim N^{-2}$ is an infrared cut-off introduced to compute the correlation function in the thermodynamic limit, and}}
the correlation time as
\beq
\tau_\text{corr}  \sim\varepsilon^{-1} \sim N^2.
\eeq

We see that for supercritical systems, both the longest time scale $\tau_\text{max}$ and the correlation time $\tau_\text{corr}$ increase as a power of the system size; the rate is even faster than the system at criticality. In fact, there are divergently many slow modes. Meanwhile, the interaction strength can undertake a range of values, as long as they are greater than a certain threshold. Conceptually, with the two-dimensional electron gas analogy, the hard threshold imposes a semi-infinite well in addition to the quadratic potential and logarithmic repulsion. Although divergently many eigenvalues cluster near the stability threshold, the repulsion among the eigenvalues indeed pushes the smallest eigenvalue further away from the stability threshold, introducing shorter timescales to the system.

Nonetheless, we cannot quite claim that the problem is solved.
First, in the supercritical phase, the correlation function does not decay as a power law. Instead, the correlation function stays at 1 for a time period $\tau_\text{corr}\sim \tau_{\rm max}$, and then decays exponentially. This means the system has a single long time scale, rather than a continuous spectrum of slow modes. Second, in order for a system to impose a hard constraint on its stability, it needs to measure its stability.  Naively, checking for stability, especially in the presence of slow modes,  requires access to infinitely long measuring times; implementing a sharp threshold may also be challenging.

\subsection{Model 3: Constraining mean-square activity}
\label{subsec:lts_soft:def}

While it can be difficult to check for stability, it is much easier to imagine checking the overall level of activity in the network.  One can even think about mechanisms that would couple indirectly to activity, such as the metabolic load.  If the total activity is larger than some target level, the system might be veering toward instability, and there could be feedback mechanisms to reduce the overall strength of connections.  Regulation of this qualitative form is known to occur in the brain, and is termed synaptic scaling~\cite{turrigiano_activity-dependent_1998, turrigiano_homeostatic_2004, abbott_synaptic_2000}; this is hypothesized to play an important role in maintaining persistent neural activities~\cite{renart_robust_2003, tetzlaff_synaptic_2013}.   
In this section,  we construct the least structured distribution $P(\bm M)$ that is consistent with a fixed mean (square) level of activity, which we can think of as a soft threshold on stability, and derive the density of eigenvalues that follow from this distribution.  
In the following section  we discuss possible mechanisms for a system to generate matrices $\bm M$, dynamically, out of this ensemble.

\subsubsection{The spectral density}

It is useful to remember that the GOE, Eq.~(\ref{GOE1}), can be seen as the maximum entropy distribution of matrices consistent with some fixed variance of the matrix elements $M_{ij}$ ~\cite{jaynes57, presse_principles_2013}.  If we want to add a constraint, we can stay within the maximum entropy framework, and in this way we isolate the generic consequences of this constraint:  we are constructing the least structured ensemble of networks that satisfies the added condition.

Recall that if we want to constrain the mean values of several functions $f_\mu (\bm M)$,  the maximum entropy distribution has the form
\begin{equation}
P(\bm M) \propto \exp\left[ - \sum_\mu g_\mu f_\mu (\bm M)\right],
\end{equation}
where the $g_\mu$'s are the Lagrange multipliers used to enforce the constraints~\cite{jaynes57, presse_principles_2013}. In our case, we are interested in the mean--square value of the individual matrix elements, and the mean--square value of the activity variables $x_i$.  Because our basic model Eqs.~(\ref{eq:lin_dyn_wn}) and (\ref{noise1}) predicts that
\begin{equation}
\mu = {1\over N} \sum_i \langle x^2_i\rangle = {1\over N} \sum_i {1\over{1-\lambda_i}},
\end{equation}
the relevant maximum entropy model becomes
\beq\label{model3}
P_\text{soft}(\bm M) \propto \exp\left[-\frac{N}{2c^2}\text{Tr} \bm M^\intercal \bm M -N\xi \sum_i \frac{1}{1-\lambda_i}\right],
\eeq
where $\xi$ is the Lagrange multiplier constraining the norm activity $\mu$. The scaling of $N \xi$ ensures a well-defined thermodynamic limit. {\RV{See Fig.~2(d) for schematics.}}

Again, this distribution is invariant to orthogonal transformation. After the integration over the rotation matrices, we have
\begin{widetext}
\beq\label{eq:soft_jointp_lambda}
P_\text{soft}(\lbrace\lambda_i \rbrace) \propto
\exp
\left[
-\frac{N}{2c^2}\sum_i\lambda_i^2
+\frac{1}{2}\sum_{j\neq k}\ln \big\lvert \lambda_j - \lambda_k \big \rvert - N\xi \sum_i \frac{1}{1-\lambda_i}
\right].
\eeq
Mathematically, this is the Boltzmann distribution for two-dimensional electron gas, with a soft wall local potential.
Luckily, the same arguments that yield the exact density of eigenvalues in the  Gaussian Orthogonal Ensemble  also work here (see Appendix~\ref{appd: solve_spectrum}), and we find
\beq\label{eq:model3_spectrum}
\rho_\text{soft}(\lambda) = \frac{1}{\pi\sqrt{(\lambda - 1 + g_0 + l)(1 - g_0 - \lambda) }}
\left[
1 + \frac{l^2}{8 c^2} + \left(
1 - g_0 - \frac{1}{2}l
\right) \frac{\lambda}{c^2}
-\frac{\lambda^2}{c^2}
+\frac{\xi}{2} 
\frac{ (2 g_0 - 2 g_0^2 + l - 2g_0 l)- (2g_0 + l)\lambda}{\sqrt{g_0(g_0+l)}(\lambda - 1)^2 }
\right]\mbox{,}
\eeq
\end{widetext}
where the gap size $g_0$ and the width of the support $l$ are solved numerically by setting the spectral density at the two ends of the support to zero.

To our surprise, unlike for GOE with hard stability constraints as in Model 2, here we find a finite gap for all $\xi > 0$ (see Appendix C for calculation). Intuitively, this is due to the soft threshold acting as a repelling force on the eigenvalues to push them away from the stability threshold. This finite gap means that there exists a maximum time scale even when the system is infinitely large. This upper limit of longest time scale depends on the Lagrange multiplier $\xi$ and the interaction strength $c$. Because the Lagrange multiplier $\xi$ is used to constrain the averaged norm activity $\mu$, the maximum time scale is set by the allowed norm of the activity $\mu$, measured in units of expected norm for independent neurons (Fig.~\ref{fig:soft_mf}(e)). As we explain below, the greater dynamic range the system leads to longer time scales.

The dependence of the gap on the Lagrange multiplier $\xi$ is shown in Fig.~\ref{fig:spec}{\RV{(c)}} at each of several fixed values of the interaction strength $c$. In the limit of $\xi \rightarrow 0$, we recover the hard wall case, where there is a phase transition at $c_\text{crit} = 1/\sqrt{2}$. This phase transition remains present for $0 < \xi  \ll 1$, with the eigenvalues close to the stability threshold pushed into the bulk spectrum (see details in Section~\ref{sec:time_scaling} and Appendix~\ref{sec:lts_appd:scaling_calc}).  For large $\xi > 1$, the entire spectrum is pushed away from the wall. A closer look at the longest time scale $\tau_\text{max}$ vs. Lagrange multiplier in Fig.~\ref{fig:soft_mf}(a) confirms that amplification of time scales occurs only when $\xi < 1$, corresponding to an amplification of mean--square activity $\mu$.

\subsubsection{The scaling of time scales in three phases}\label{sec:time_scaling}

\begin{figure}[ht!]
\includegraphics[scale=1]{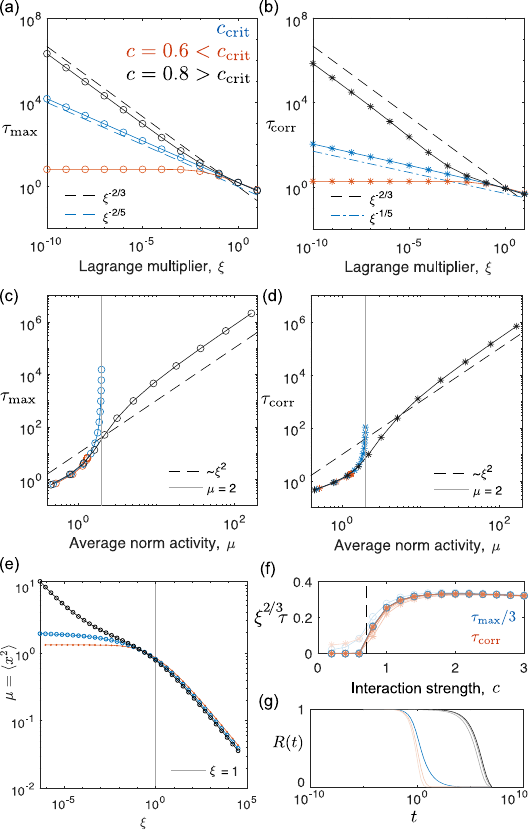}
\caption{Mean field results for Model 3 {\RV{constraining the mean-square activity. (a) The longest time scale $\tau_\text{max}$ vs.~the Lagrange multiplier $\xi$ which constrains the mean-square activity, for interaction strength $c = 0.6$ (subcritical, \textit{red}), $c = c_\text{crit} = 1/\sqrt{2}$ (critical, \textit{blue}), $c = 0.8$ (supercritical, \textit{black}). The dashed lines bound the scaling behavior at small $\xi$ (see Appendix D). (b) Same as (a) but for the correlation time scale $\tau_\text{corr}$. (c, d) Same as (a, b), but plotting against the average norm activity $\mu$. At $c = c_\text{crit}$, the norm activity asymptotically approaches $\mu = 2$. (e) Average norm activity $\mu$ as a function of the Lagrange multiplier $\xi$.  (f) The exact scaling and the amplification of time constant depending on whether the system is subcritical (with interaction strength $c < 1/\sqrt{2}$), critical ($c = 1/\sqrt{2}$), or supercritical. The different shade indicates different values of the Lagrange multiplier $\xi$: from lightest to darkest, $\xi  = 10^{-1}, 10^{-4}, 10^{-7}, 10^{-10}$, respectively (see Fig.~4(e) legend).
%The longest time scale $\tau_\text{max}$ (panel (a)) and the correlation time scale $\tau_\text{corr}$ (panel (b)) increase with different scaling as the Lagrange multiplier $\xi$ decreases, corresponding to an increasing value for the constrained averaged norm activity, $\mu$ (panel (c,d,e)). The exact scaling and the amplification of time constant depending on whether the system is subcritical (with interaction strength $c < 1/\sqrt{2}$), critical ($c = 1/\sqrt{2}$), or supercritical (panel (f)). For panel (f), the different shade indicates different values of the Lagrange multiplier $\xi$: from lightest to darkest, $\xi  = 10^{-1}, 10^{-4}, 10^{-7}, 10^{-10}$, respectively (see Fig.~4(e) legend). %{\RV{\sout{Despite the supercritical phase exhibit long time scales, the autocorrelation function decays as a power law only at $c = c_\text{crit}$, and as exponential for other values of interaction strength (panel (g)). For this panel, $\xi = 10^{-10}$, red curves are for $c = 0.2$, $c = 0.6$, blue curve is at $c = c_\text{crit}$, and black curves are at $c = 1$ and $c = 3$.}
(g) Normalized correlation function $R(t)$ computed at $\xi = 10^{-10}$. For each curve from left to right, the interaction strength is $c = 0.2$, $c = 0.6$ (\textit{red}), $c = c_\text{crit}$ (\textit{blue}), $c = 1$, $c = 2$ and $c = 3$ (\textit{black}), respectively.}}
}\label{fig:soft_mf}
\end{figure}

We now discuss the dependence of time scales on the interaction strength $c$. In contrast to the ensemble with a hard stability threshold, we find a finite gap  for all values of interaction strength $c$, but the scaling of time scales vs. the Lagrange multiplier (and hence the mean--square activity) is different in the different phases. Details of the calculation can be found in Appendix D.

For the subcritical and critical phases, the results are as expected from the  hard wall case. In the subcritical phase, as $\xi \rightarrow 0$, the spectral distribution converges smoothly to the familiar semicircle. The time scales and the mean--square activity both approach constants {\RV{(Fig.~\ref{fig:soft_mf}(a,b,e), red curves)}}. On the other hand, when $c = c_\text{crit}$, we find that the longest time scale grows as $\tau_\text{max} \sim \xi^{-2/5}$, and the correlation time scale $\tau_\text{corr} \sim \xi^{-1/5}$, i.e. both time scales can be large if $\xi$ is small enough {\RV{(Fig.~\ref{fig:soft_mf}(a, b), blue curves)}}. Meanwhile, the norm activity $\mu$ asymptotically approaches a constant value of $\mu = 2$ {\RV{(Fig.~\ref{fig:soft_mf}(c, d), blue curves)}}. This suggests that if a system is poised at criticality, then the system can exhibit long time scales, even when the dynamic range of individual components is well controlled. The autocorrelation function exhibits a power law-like decay, as expected; see the blue curve in Fig.~\ref{fig:soft_mf}(g). 

The most interesting case is the supercritical phase, where the interaction strength $c > c_\text{crit} = 1/\sqrt{2}$. As $\xi \rightarrow 0$, the spectrum does not converge to the spectrum with the hard constraint. 
We find that both time scales and the norm activity increase as power laws of the Lagrange multiplier $\xi$ (Fig.~\ref{fig:soft_mf}(a{\RV{-d), black curves}}), with $\tau_\text{max} \approx 3\tau_\text{corr} \sim \xi^{-2/3}$, and $\mu \sim \xi^{-1/3}$.
 This implies that the time scales grow as a power of the allowed dynamic range of the system, although not with the size of the system.  The question of whether the resulting time scales are ``long'' then becomes more subtle.  Quantitatively, we see from  Fig.~\ref{fig:soft_mf}(c),  that if the system has an allowed dynamic range just $10\times$  that of independent neurons, the system can generate  time scales $\tau_\text{max}$  almost $10^4\times$ longer than the time scale of isolated neurons. Similar amplification of the correlation time can been seen in Fig.~\ref{fig:soft_mf}(d).

Interestingly, once the system is in the supercritical phase, the ratio of amplification has only a small dependence on the interaction strength $c$ (Fig.~\ref{fig:soft_mf}(f)). Intuitively, an increasing interaction strength $c$ implies that without constraints more modes will be unstable, while with constraints more modes concentrate near the stability threshold, but the entire support of the spectrum also expands (see Fig.~2(d) and Appendix A Eq.~(A7)), so the density of slow modes and their distance to the stability threshold remain similar. 
This is perhaps another indication for long time scales without fine tuning when the system uses its norm activity to regulate its connection matrix. 

We note that, although both the critical phase and the supercritical phase can reach time scales that are as long as the dynamic range allows, there are significant differences between the two phases. One difference is that in the critical phase, locally  the dynamic range for each neuron can remain finite, while for the supercritical phase,  the variance of activity for individual neurons can be much greater. Moreover, as shown by Fig.~\ref{fig:soft_mf}(g), systems in the supercritical phase are dominated by a single slow mode, rather than by a continuous spectrum of slow modes. While the autocorrelation function decays as a power law in the critical phase, in the supercritical phase, it holds at $R(t) = 1$ for a much longer time compared to the subcritical case, but then decays exponentially.  While a single long time scale can be achieved without fine tuning, it seems that a continuous spectrum of long time scales is much more challenging.

\subsubsection{Finite Size Effects}
\label{subsubsec:lts_soft:fss}

If we want these ideas to be relevant to real biological systems, we need to understand what happens at  finite $N$.  We investigate this numerically using direct Monte-Carlo sampling of the (joint) eigenvalue distribution.  As the system size grows, the time scales $\tau_\text{max}$ and $\tau_\text{corr}$ also increase, up to the upper limit given by the mean field results; see Fig.~\ref{fig:soft_fss}{\RV{(a, b)}}. {\RV{The relation between the timescales and the average norm activity does not depend on the system size (Fig.~\ref{fig:soft_fss}(c, d)).}} {\RV{Interestingly,}} the scaling exponent $\alpha$ for the gap difference,  
\begin{equation}
\Delta g_0(N) \equiv g_0(N) - g_0 \sim N^\alpha,
\end{equation}
depends on the Lagrange multiplier $\xi$ (Fig.~\ref{fig:soft_fss}(e,f)). This is supported by the analytical argument that the scaling interpolates between two limiting cases: for small $\xi$, the gap scales as $\Delta g_0 \sim N^{-2}$, with the same scaling factor as in the case with 
the hard threshold; for large $\xi$, the gap scales as $\Delta g_0 \sim N^{-2/3}$, which has the same scaling factor as in 
the Gaussian ensemble without any additional constraint.  In any case, thousands of neurons will be well described by the $N\rightarrow\infty$ limit.

\begin{figure}[b]
\includegraphics[scale=1]{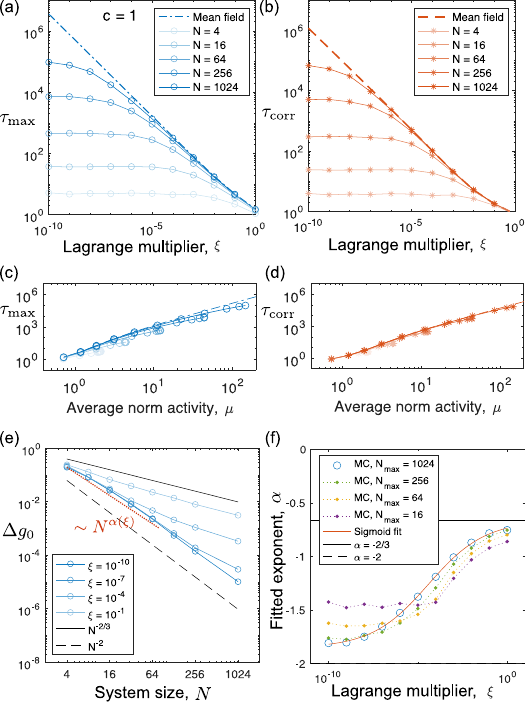}
\caption{Finite size effects on the time scales $\tau_\text{max}$ and $\tau_\text{corr}$ for Model 3, super-critical phase, with interaction strength $c = 1$. (a) The longest time scale $\tau_\text{max}$ as a function of the Lagrange multiplier $\xi$.
Results from direct Monte Carlo sampling of the eigenvalue distribution are plotted together with the mean field results. 
(b) Same as panel (a) but for the correlation time scale $\tau_\text{corr}$. (c) {\RV{The longest timescale}} $\tau_\text{max}$ as a function of the norm activity $\mu$. The difference among different-sized systems is small. (d) Same as (c) but for the correlation time scale $\tau_\text{corr}$.
{\RV{(e) The gap difference $\Delta g_0 (N) \equiv g_0(N) - g_0$ vs.~system size $N$. 
(f) }}
%The fitted exponent $\alpha$ from panel (e), defined by $\Delta g_0(N) \sim N^\alpha$, as a function of the Lagrange multiplier $\xi$ which is used to constrain the mean-square activity. }}
 The exponent $\alpha$ that explains the convergence of $\tau_\text{max}$ when system size increases depends on the Lagrange multiplier $\xi$, interpolating between $\alpha = 2$ and $\alpha = 2/3$. {\RV{The \textit{red} curve shows a sigmoidal fit.}}
}
\label{fig:soft_fss}
\end{figure}

\subsubsection{Distribution of matrix elements}
\label{subsubsec:lts_soft:mii}

Now that we have examples of network ensembles that exhibit long time scales, we need to go back and check what these ensembles predict for the distribution of individual matrix elements. In particular, because we did not constrain the self interaction $M_{ii}$ to be 0, we want to check whether the long time scales emerge as a collective behavior of the network, or trivially from an effective increase of the intrinsic time scales for individual neurons. 
A similar question arises in real networks, where there have been  debates about the importance of feedback within single neurons vs. the network dynamics in maintaining long time scales; see Ref.~\cite{major_persistent_2004} for a review.  We confirm that, at least in our models, this is not an issue: the constraint on the norm activity, in fact, pushes the average eigenvalues to be negative, and hence the effective self interaction for individual neurons actually leads to a shorter intrinsic time scale. We can   impose additional constraints on the distribution of matrix elements, such that $\langle M_{ii}\rangle = 0$; for this ensemble, we can again solve for the spectral distribution, and we find that scaling behaviors described above do not change (figures not shown).

\begin{center}
\begin{figure*}
\includegraphics[scale=1]{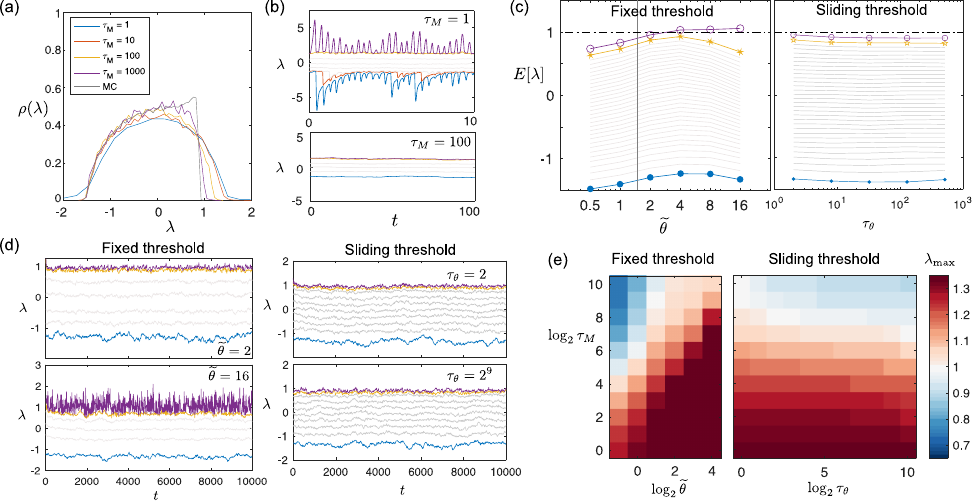}
\caption{
%The ensemble of connection matrices with long time scales can be achieved dynamically by implementing a Langevin dynamics with a timescale of $\tau_M$ to update the connection strength, either with a fixed threshold $\widetilde{\theta}$ or a sliding threshold with a timescale $\tau_\theta$ constraining the total norm of neural activities. 
(a) The spectral distribution for connection matrix drawn from the dynamics with fixed threshold approaches the static distribution as the time constant $\tau_M$ increases. 
(b)
{\RV{Dynamics of representative eigenvalues with the optimal fixed threshold $\widetilde{\theta} = \theta^*$, with $N = 64$, and time constant $\tau_M = 1$ and 100. }}
% \textcolor{red}{[Change this!!! Representative eigenvalues for $N = 64$.]} 
If the connection matrix updates too fast (\textit{top}), i.e. $\tau_M$ is too small, the system exhibits quasi-periodic oscillations, and does not reach a steady state distribution. %In contrast, long $\tau_M$ leads to the adiabatic approximation for the steady state distribution. 
(c) The expected eigenvalues for the dynamics with fixed threshold $\widetilde{\theta}$, and for the ones with sliding threshold, with $N = 32$, $c = 1$, $\xi = 2^{-5}$ and $\tau_M = 1000$. (d) Example traces for the eigenvalues vs.~time, for fixed thresholds at different threshold values (\textit{left}), and for sliding thresholds at different updating timescales $\tau_\theta$. (e) {\RV{Heatmap for expected $\lambda_{\max}$ as a function of dynamical model parameters, i.e. the time constant $\tau_M$, the fixed threshold $\widetilde{\theta}$ (\textit{left}), and the sliding threshold time scale $\tau_\theta$ (\textit{right}).}} 
%Average $\lambda_{\max}$ as a function of model parameters for dynamics with a fixed threshold (left) and a sliding threshold (right). 
At equal $\tau_M$, the dynamics with a sliding threshold has a much larger range of parameter $\tau_\theta$ for the system to exhibit slow modes ($\lambda_\text{max} \lessapprox 1$), while the dynamics with fixed threshold requires fine-tuning of its parameter $\widetilde{\theta}$.
}
\label{fig:dyn}
\end{figure*}
\end{center}

\subsection{Dynamic tuning}
\label{subsubsec:soft:tune_dyn}
So far, we have established that a distribution constraining the norm activity can lead generically to long time scales, but we haven't really found a mechanism for implementing this idea.  But if we can write the distribution of the interaction matrices $\bm M$ in the form a Boltzmann distribution, we know that we can sample this distribution by allowing the matrix elements be dynamical variables undergoing Brownian motion in the effective potential.  We will see that this sort of dynamics is closely related to previous work on self--tuning to criticality  ~\cite{magnasco_self-tuned_2009}, and we can interpret the dynamics as implementing familiar ideas about synaptic dynamics, such as Hebbian learning and metaplasticity.

We can rewrite our model in Eq.~(\ref{model3}) as
\beqn
P_\text{soft}(\bm M) &=& {1\over Z}\exp\left[-\frac{N}{2c^2}\text{Tr} \bm M^\intercal \bm M -N\xi \sum_i \frac{1}{1-\lambda_i}\right] \nonumber\\
&=& {1\over Z}\exp\left[- V(\bm M)/T\right],\\
V(\bm M) &=& \frac{1}{2}\text{Tr} \bm M^\intercal \bm M + c^2\xi \text{Tr}(1-\bm M)^{-1},
\eeqn
with a temperature $T = c^2 /N$.   The matrix $\bm M$ will be drawn from the distribution $P(\bm M)$, as $\bm M$ itself performs Brownian motion or Langevin dynamics in the potential $V(\bm M)$:
\beq
\begin{split}
\tau_M \dot{\bm M} &  = -\frac{\partial V(\bm M)}{\partial \bm M} + \bm \zeta(t) \\
& = -\bm M - c^2 \xi (\mathbf{1} - \bm M)^{-2} + \bm \zeta(t),
\end{split}
\eeq
where the noise has zero mean, $\langle \bm \zeta \rangle = \bm 0$, and is independent for each matrix element,
\beq
 \langle \zeta_{ij}(t)\zeta_{kl}(t') \rangle= 2T\tau_M \delta_{ik}\delta_{jl}\delta(t-t').
\eeq

It is useful to remember that, in steady state, our dynamical model for the $\{x_i\}$, Eqs.~(\ref{eq:lin_dyn_wn}) and (\ref{noise1}), predicts that
\beq
\langle x_i x_j \rangle = [(\mathbf{1} -\bm M)^{-1} ]_{ij} .
\eeq
This means that we can rewrite the Langevin dynamics of $\bm M$, element by element, as
\beq
\begin{split}
\tau_M \dot{M}_{ij} & =  -M_{ij} - c^2 \xi [(\mathbf{1} - \bm M)^{-2}]_{ij}+ \zeta_{ij}(t) \nonumber\\
& =  - M_{ij}  - c^2 \xi \langle x_i x_k \rangle \langle x_k x_j   \rangle + \zeta_{ij}(t).
\end{split}
\eeq
Because the $x_i$ are Gaussian, we have
\beq
\begin{split}
\sum_k \langle x_i x_k \rangle \langle x_k x_j   \rangle & = {1\over 2} \sum_k \left( \langle x_i x_k x_k x_j \rangle - \langle x_i x_j\rangle \langle x_k x_k\rangle\right) \\
& = {1\over 2} {\bigg\langle} x_i x_j  \left(\sum_k x_k^2 - \sum_k\langle x_k^2\rangle\right){\bigg\rangle}.
\end{split}
\eeq

We now imagine that the dynamics of $\bm M$ is sufficiently slow that we can replace averages by instantaneous values, and let the dynamics of $\bm M$ do the averaging for us.  In this approximation we have
\begin{equation}
\tau_M \dot{M}_{ij}  =  -M_{ij} - {1\over 2}c^2 \xi x_i x_j \left(\sum_k x_k^2 - \theta \right) + \zeta_{ij}(t) ,
\label{langevinM}
\end{equation}
where the threshold $\theta = \sum_k \langle x_k^2\rangle$.

The terms in this Langevin dynamics have a natural neuroscience interpretation.  First, the connection strength decays with an overall time constant $\tau_M$.  Second, the synaptic connection $M_{ij}$ is driven by the correlation between pre-- and post--synaptic activity, $\sim x_i x_j$, as in Hebbian learning~\cite{hebb_organization_1949, magee_synaptic_2020}.  The response to correlations is modulated depending on whether the global neural activity is greater or less than a threshold value. 
If the network is highly active, then the connection between neurons with correlated activity will decrease, i.e. the dynamics are anti-Hebbian; if the overall network is quiet, the dynamics are Hebbian.  

We still have the problem of setting the threshold $\theta$. {\RV{The first option is to set the threshold fixed using a constant, $\widetilde{\theta}$.}}  Ideally, for the dynamics to generate samples out of the correct $P(\bm{M})$, we need
\beq
\widetilde{\theta} = \theta^* \equiv \langle x^\intercal x \rangle_\text{s.s.} = N\mu(c,\xi),
\eeq
where as above $\mu$ is the mean activity whose value is enforced by the Lagrange multiplier $\xi$.   This means that the fixed threshold $\widetilde{\theta}$ needs to be tuned in relation to $\xi$, and it is challenging to have a mechanism that does this directly, and just pushes the fine tuning problem back one step.  Importantly, if  $\widetilde{\theta} = \theta^*$ then the steady state spectral density of the connection matrix approaches the desired equilibrium distribution as the update time constant increases (Fig.~\ref{fig:dyn}{\RV{(a)}}).  {\RV{ As expected, if the connection matrix updates too fast, i.e. $\tau_M$ is too small, the system exhibits quasi-periodic oscillations, and does not reach a steady state distribution. In contrast, long $\tau_M$ leads to the adiabatic approximation for the steady state distribution (Fig.~\ref{fig:dyn}(b)).}}
Nonetheless, if  $\widetilde{\theta}$ deviates from $\theta^*$, then the steady distribution does not have slow modes.
As shown in Fig.~\ref{fig:dyn}{\RV{(c)}}, if the threshold is too small, then the entire spectrum is shifted away from the stability threshold, and the system no longer exhibits long time scales; if the threshold is too large, then the largest eigenvalue oscillates around the stability threshold $\lambda = 1$, and a typical connection matrix drawn from the steady state distribution is unstable {\RV{(Fig.~\ref{fig:dyn}(d))}}. 

But we can once again relieve the fine tuning problem by promoting $\theta$ to a dynamical variable,
\beq
\tau_\theta \dot{\theta} = \sum_k x_k^2 - \theta,
\eeq
which we can think of as a sliding threshold in the spirit of models for metaplasticity~\cite{bcm_1982}.  We pay the price of introducing yet another new time scale, $\tau_\theta$, but in Fig.~\ref{fig:dyn}(d,e) we see that this can vary over at least three orders of magnitude without significantly changing the spectral density of the eigenvalues.

To see whether the sliding threshold really works, we can compare results where $\theta$ is fixed to those where it changes dynamically; we follow the mean value of $\lambda_\text{max}$ as an indicator of system performance.  We  choose parameters in the supercritical phase, specifically $c = 1$ and $\xi = 2^{-5}$, and study a system with  $N = 32$.   Figure~\ref{fig:dyn}(e) shows that with fixed threshold, even in the adiabatic limit where $\tau_M \gg 1$, there is only a measure-zero range for the fixed threshold $\widetilde{\theta}$ such that $\lambda_\text{max}$ is very close to, but smaller than $1${\RV{, and hence requires fine-tuning of its parameters}}. In contrast, for the dynamics with sliding threshold, at $\tau_M \gg 1$ there is a large range of values for the time constant $\tau_\theta$ such that the system hovers just below instability, generating a long time scale.
%{\RV{[Check. this is what we had previously in the caption] 
%Average $\lambda_{\max}$ as a function of model parameters for dynamics with a fixed threshold (left) and a sliding threshold (right). At equal $\tau_M$, the dynamics with a sliding threshold has a much larger range of parameter $\tau_\theta$ for the system to exhibit slow modes ($\lambda_\text{max} \lessapprox 1$), while the dynamics with fixed threshold requires fine-tuning of its parameter $\tilde{\theta}$}}

\section{Discussion}

Living systems can generate behaviors on time scales that are much longer than the typical time scales of  their component parts, and in some cases correlations in these behaviors decay as an approximate power--law, suggesting a continuous spectrum of slow modes. In order to understand how surprised we should be by these behaviors, it is essential to ask whether there exist biologically plausible dynamical systems that can generate these long time scales. We have constructed a mechanism for
a broad class of models for interacting living systems--symmetric linear dynamical models--to reach long time scales with the least stringent fine-tuning condition: when the interaction strength of the connection matrix is large enough, imposing a global constraint on the stability of the system leads to divergently many slow modes. To impose a biologically plausible mechanism for living systems, we constrain the averaged norm activity as a proxy for global stability; in this case, the time scales for the slow modes are set by the allowed dynamic range of the system. 

Further, we showed that ensembles of interaction matrix $\bm{M}$ with long time scales can be achieved dynamically by updating $\bm{M}$ with a sliding threshold for the norm activity, a mechanism that resembles homeostatic synaptic plasticity in neuronal networks, particularly through mechanisms such as synaptic scaling and switching between Hebbian and anti-Hebbian learning by thresholding the average neural firing rate~\cite{turrigiano_activity-dependent_1998, turrigiano_homeostatic_2004}.
Biologically, it has been suggested that global modulation of synaptic strengths across a spatially extended network can be effected by activity-dependent secretion of molecules such as brain-derived neurotrophic factor, tumor necrosis factor-$\alpha$, and retinoic acid~\cite{homeostasis_mechanism_review_2010, turrigiano2012homeostatic}. From the modeling perspective, we can compare the Langevin dynamics of our system to the BCM theory of metaplasticity in neural networks, in that both models involve a combination of Hebbian learning and a threshold on the neural activity~\cite{bcm_1982}.  
Both dynamics have more controlled dynamics when the activity threshold is chosen to be sliding compared to fixed, but
 there are two key differences. First, the BCM theory imposes a threshold on the activity of locally connected neurons, while here the threshold is on the overall neural activity. Second, the BCM dynamics is Hebbian when the postsynaptic activities are larger than the threshold, and anti--Hebbian otherwise, which is the opposite of the dynamics for our system. It is interesting that in some other models for homeostasis, plasticity requires the activity detection mechanism to be fast ($\tau_\theta/\tau_M \ll 1$) for the system to be stable~\cite{zenke_synaptic_2013, zenke_temporal_2017}, which we do not observe for our system.

Importantly, the slow modes achieved through constraining norm activity typically lead to exponentially decaying correlations; only when the interaction strength of the matrix is at a critical value do we find power-law decays.  This suggests that a continuous range of slow modes is more difficult to achieve than a  single long time scale. A natural follow-up question is whether there exist mechanisms which can tune the system to criticality in a self-organized way, for example by coupling the interaction strength to the averaged norm activity.

To achieve long time scales a high dimensional dynamical system requires some degree of fine tuning: from the most to the least stringent, examples include setting individual elements of the connection matrix by coupling the system to the environment~\cite{major_plasticity_2004}, or choosing a particular network architecture such as a ring or grid network~\cite{goldman_memory_2009, burak_fundamental_2012, shi2023spatial}. 
All of those network designs require tuning of either the structure (i.e. whether certain neurons are connected) or the interaction strength (i.e. how strong is the connection). Comparing to other models with a focus on tuning the network structure, our work focuses on the distribution of interaction strength. Starting from all-to-all connected networks allow us to be as general as possible, which is especially useful when one considers a dynamical process that leads to such functional networks. Possible future work includes pruning the network where the interaction strength is small, which will allow a comparison between the emergent network structure and other works which start with a fixed structure.

Both for simplicity and to understand the most basic picture, we have been focusing on linear networks with symmetric connections. Realistically, many biological networks are asymmetric, which gives rise to more complex and even chaotic dynamics~\cite{sompolinsky_chaos_1988, vreeswijk_chaos_1996}. In the asymmetric case, the eigenvalue spectrum for the Gaussian ensemble is well known (uniform distribution inside a unit circle)~\cite{ginibre_statistical_1965, forrester_eigenvalue_2007}, but a similar global constraint on the norm activity leads to a dependence on the overlap among the left and right eigenvectors (see Appendix E and Ref.~\cite{chalker_eigenvector_1998, mehlig_statistical_2000}). In particular, the matrix distribution can no longer be separated into the product of eigenvalues and eigenvectors, and it is difficult to solve for the spectral distribution analytically. Two new features emerge in the asymmetric case.  First, the time scales given by real eigenvalues vs. complex eigenvalues may be different, leading to more (or less) dominant oscillatory slow modes in large systems~\cite{akemann_integrable_2007}. Second, asymmetric connection matrices can lead to complicated transient activity when the system is perturbed, with the time scales mostly dominated by the eigenvector overlaps, and can be very different from the time scales given by the eigenvalues~\cite{grela_what_2017}. In the limit of strong asymmetry, the network is organized into a feed-forward line, which with fine tuning of interactions, information can be held for a time that is extensive in system size; see examples in Refs.~\cite{goldman_memory_2009} and~\cite{ganguli_memory_2008}. It will be interesting to check whether systems can store information in these transients without fine-tuning the structure of the network.

While we compare this dynamical updating mechanism to synaptic updating mechanisms in neuroscience, note that the methods we propose is conceptual and can be applied generally to models of interacting systems, and is not limited to neurons.
The system we study can be extended to consider more specific ensembles for particular biological systems. For example, real neural networks have inhibitory and excitatory neurons, so that elements belonging to the same column need to share the same sign, and the resulting spectral distribution has been shown to differ from the unit sphere~\cite{rajan_eigenvalue_2006}; more generally recent work explores how structured connectivity can lead to new but still universal dynamics~\cite{tarnowski_universal_2019}. Another limitation of our work is that it only considers linear dynamics, or only dynamical systems where all fixed points to be equally likely. In contrast,  non-linear dynamics such as the Lotka-Volterra model in ecology~\cite{biroli_marginally_2018}, discrete spiking dynamics in neuronal nets~\cite{harish_hansel_2015}, and the gating neural network in machine learning~\cite{can_gating_2020, krishnamurthy_theory_2020} have been shown to drive systems to non-generic, marginally stable fixed points, around which there exists an extensive number of slow directions for the dynamics. In summary, we believe the issue of whether a continuum of slow modes can arise generically in neural networks remains open, but we hope that our study of very simple models has helped to clarify this question.

\section*{Code availability}
Custom codes were developed in C to implement standard Monte Carlo sampling methods and stochastic dynamics. The codes are available at https://doi.org/10.5281/zenodo.12103680.

\begin{acknowledgments}
We thank Hanrong Chen and Kamesh Krishnamurthy for fruitful discussions. This work was supported in part by the National Science Foundation, through the Center for the Physics of Biological Function (PHY-1734030) and Grant PHY--1607612, and by the National Institutes of Health through Grants NS104889 and R01EB026943--01.
\end{acknowledgments}

\bibliographystyle{unsrt}
\bibliography{../ref/thesis.bib}

\pagebreak
\newpage
\appendix
%\onecolumngrid

\hspace{1cm}
\section{Spectral distribution for the Gaussian Orthogonal Ensemble with eigenvalue constraints}
\label{appd: solve_spectrum}
For completeness, we sketch here the %well-known 
derivation of the spectral density for random matrices drawn from the Gaussian Orthogonal Ensemble~\cite{wigner_on_1951, dyson_threefold_1962, dyson_brownianmotion_1962}, subject to constraints on the eigenvalues. Excellent pedagogical discussions can be found in Refs~\cite{marino_number_2016, livan_introduction_2018}.   These 
methods allow us to derive the spectral densities in all 
cases that we consider in the main text.

Let $\bm{M}$ be a matrix with size $N \times N$. Assume $\bm{M}$ is real and symmetric ($M_{ij} = M_{ji}$), and that the individual elements of the $\bm{M}$ matrix are independent gaussian random numbers, 
\beqs
\begin{split}
M_{ii} & \sim \mathcal{N}(0, c^2/N),\\
M_{ij}\vert_{i\neq j} & \sim \mathcal{N}(0, c^2/2N).
\end{split}
\eeqs
Here, $c$ is the interaction strength.
This is the Gaussian Orthogonal Ensemble (GOE). Together with its complex and quaternion counterparts, the Gaussian Ensembles are the only random matrix ensembles that both have independent entries and are invariant under orthogonal (unitary, symplectic) transformations, which is more obvious when we write the probability distribution of $\bm{M}$ in terms of its trace:
\beq
P_\text{GOE}(\bm{M}) \propto \exp\left( -
\frac{N}{2c^2}\Tr \bm{M}^\intercal \bm{M}
\right).
\eeq

More generally, we consider the ensemble of random matrices with additional constraints on the eigenvalues, i.e. where the probably of a given matrix $\bm{M}$ can be written as
\beq
P(\bm{M}) \propto
\exp 
\left( 
 - N\sum_i f(\lambda_i)
-\frac{N}{2c^2}\Tr  \bm{M}^\intercal \bm{M}
 \right).
\eeq
Symmetric matrices can be diagonalized by orthogonal transformations,
\beqs
\bm{M} = \bm{O}^\intercal \bm{\Lambda} \bm{O},
\eeqs
where the matrix $\bm{O}$ is constructed out of the eigenvectors of $\bm{M}$ and the matrix $\bm{\Lambda}$ is diagonal with elements given by the eigenvalues $\{\lambda_i\}$.  Because $P(\bm{M})$  is invariant to orthogonal transformations of $\bm{M}$, it is natural to integrate over these transformations and obtain the joint distribution of eigenvalues.  To do this we need the Jacobian, {\RV{which is given by}} the Vandermonde determinant,
\beqs
d\bm{M} {\RV{= \prod_{i \leq j} dM_{ij}}} 
= \prod_{i<j}\lvert \lambda_i -\lambda_j \vert d\mu(\bm{O}) \prod_{i=1}^N d\lambda_i ,
\eeqs
where $d\mu(\bm{O})$ is the Haar measure of the orthogonal group under its own action.  Now we can integrate over the matrices $\bm{O}$, or equivalently over the eigenvectors, to obtain  
\begin{widetext}
\beqs
\begin{split}
P(\lbrace\lambda_i\rbrace)\prod_{i=1}^N d\lambda_i  &= \int d\mu(\bm{O}) \prod_{i<j}\lvert \lambda_i -\lambda_j \vert  \,P(\bm{M}) \prod_{i=1}^N d\lambda_i, \\
P(\lbrace\lambda_i\rbrace) &\propto \exp 
\left(
-N\sum_i f(\lambda_i)
-\frac{N}{2c^2}\sum_i\lambda_i^2 + \frac{1}{2}\sum_{j\neq k} \ln \big\lvert \lambda_j - \lambda_k \big\rvert
\right) \\
&\propto \exp 
\left(
-N\sum_i u(\lambda_i)
+ \frac{1}{2}\sum_{j\neq k} \ln \big\lvert \lambda_j - \lambda_k \big\rvert
\right),
\end{split}
\eeqs
\end{widetext}
where we have defined
\beqs
u(\lambda) = f(\lambda) + \frac{1}{2c^2}\lambda^2.
\eeqs
Mathematically, the distribution $P(\{\lambda_i\})$ is equivalent to the Boltzmann distribution of a two-dimensional electron gas confined to one dimension, where the electrons are also experiencing a local potential with strength $u(\lambda_i)$. 
In addition, each pair of the eigenvalues repel each other with logarithmic strength; this term comes from the Vandermonde determinant, and gives all the universal features for Gaussian ensembles. 

Now we would like to solve for the spectral density. In mean-field theory, which for these problems becomes exact in the thermodynamic limit $N\rightarrow\infty$, we can replace sums over eigenvalues by integrals over the spectral density,
\beqs
\rho(\lambda) = \frac{1}{N}\sum_i \delta(\lambda - \lambda_i) .
\eeqs
Then the eigenvalue distribution can be approximated by\footnote{The double integral over the log difference needs to be corrected by the self-interaction. Luckily, these terms, after summation, are of order $N$, which is small compared to other terms with order $N^2$.}
\beqs
P(\rho(\lambda)) \propto 
\exp \left[ - N^2 S[\rho(\lambda)] \right],
\eeqs
where
\beqs
S\left[\rho(\lambda))\right] =  
\int d\lambda \rho(\lambda) u(\lambda)
- \frac{1}{2} \int d\lambda d\lambda'
\rho(\lambda) \rho(\lambda')
\ln \big\lvert \lambda - \lambda' \big\rvert.
\eeqs
Because $N$ is large, the probability distribution is dominated by the saddle point, $\rho^*$, such that
\beqs
\frac{\delta \tilde{S}}{\delta\rho}\Big\vert_{\rho = \rho^*} = 0,
\eeqs
where
\beqs
\tilde{S} = S + \kappa \int d\lambda \rho(\lambda) 
\eeqs
has a term with the Lagrange multiplier $\kappa$ to enforce the normalization of the density. Then, the spectral distribution satisfies
\beqs
u(\lambda) - \int d\lambda'  \rho^*(\lambda') \ln \big \lvert\lambda - \lambda' \big \rvert = -\kappa.
\eeqs

To eliminate $\kappa$ we can take a derivative with respect to $\lambda$, which gives us
\beq
\label{eq:hilbert_trans_general}
g(\lambda) \equiv \frac{d u(\lambda)}{d\lambda}
= \text{Pr}\int\frac{d\lambda' \rho(\lambda')}{\lambda - \lambda'}\,\mbox{,}
\eeq
with the integral defined by its Cauchy principal value.
Two methods are common in solving  equations of this form. One is the resolvent method, which we will not discuss in detail; see Ref~\cite{livan_introduction_2018}.  The other is the Tricomi solution~\cite{tricomi_integral_1957}, which states that for smooth enough $g(\lambda)$, the solution of  Eq (\ref{eq:hilbert_trans_general}) for the density $\rho(\lambda)$ is 
\begin{widetext}
\beq 
\label{eq:tricomi_sol}
 \rho(\lambda)  = 
 \frac{1}{\pi \sqrt{\lambda - a}\sqrt{b-\lambda}} 
\left[
C - \frac{1}{\pi}
\text{Pr}\int_a^b d\lambda'
\frac{\sqrt{\lambda' - a}\sqrt{b-\lambda'}}{\lambda - \lambda'} g(\lambda')
\right]\mbox{,}
\eeq
\end{widetext}
where $a$ and $b$ are the edges of the support, and
\beq
C = \int_a^b \rho(\lambda) d\lambda.
\eeq
If the distribution has a single region of support, then $C = 1$. If the distribution has more than one region of support, then we need to solve with Tricomi's solution separately for each support, and the normalization  changes accordingly. In general, solving the equation reduces to finding the edges of the support. 

We now apply the general solution (Eq.~(\ref{eq:tricomi_sol})) to the three models discussed in the manuscript to obtain the corresponding spectral distributions.

\underline{\textbf{Model 1, GOE:}} For the Gaussian Orthogonal Ensemble without any additional constraints on the eigenvalues, we have $f(\lambda) = 0$. {\RV{This means the local potential $u(\lambda) = \lambda^2 / 2c^2$ is symmetric, and hence the resulting eigenvalue distribution is invariant for $\lambda \rightarrow -\lambda$, and hence the edges of the support satisfies $a  = -b$. Substituting $g(\lambda) = \lambda/c^2$ into Tricomi's solution, we have the integral}}
\beqs
\frac{1}{\pi}\text{Pr}\int_{-b}^{b} d\lambda' \frac{\sqrt{\lambda' - b} \sqrt{b - \lambda'}}{\lambda-\lambda'}\lambda' = \lambda^2 - \frac{b^2}{2}.
\eeqs
We expect the density to fall to zero at the edges of the support, rather than having a jump.   Thus, we impose $\rho(a) = \rho(b) = 0$, which sets $b = \sqrt{2}c$, and the spectral density becomes
\beqs
\rho_\text{GOE}(\lambda) = \frac{1}{\pi c}\sqrt{2 - \frac{\lambda^2}{c^2}}.
\eeqs
This is Wigner's semicircle law.

\underline{\textbf{Model 2, GOE with hard threshold:}}
What is the spectral distribution for random matrices drawn from the GOE, but with all eigenvalues constrained to be less than a given threshold $\lambda_\text{thres} = 1$? This problem was first solved by Dean and Majumbda~\cite{dean_large_2006, dean_extreme_2008, majumdar_top_2014}, by recognizing that the eigenvalues remain to satisfy Eq.~(\ref{eq:tricomi_sol}) with the same $g(\lambda) = \lambda$ as in the GOE case, while the conditions for the edges of the support become different: namely, the two edges of the support are no longer symmetric with respect to $\lambda = 0$. By {\RV{setting the right edge of the support $b = 1$ and using $\rho(a) = 0$}}, one can solve for the spectral distribution and find if $c > c_\text{crit} = 1/\sqrt{2}$,
\beqs
\rho_\text{hard}(\lambda) = \frac{1}{c^2}\frac{\sqrt{\lambda + l^* - 1}}{2\pi \sqrt{1 - \lambda}} \left(
l^* - 2 \lambda
\right),
\eeqs
where the width of the support is
\beqs
l^* = \frac{2}{3}\left(
1 + \sqrt{1 + 6 c^2}
\right).
\eeqs
Intuitively, one can think about the eigenvalues with hard threshold using the analogy of the two dimensional electrons confined by a wall, or a semi-well potential.

\underline{\textbf{Model 3, GOE with soft threshold:}} In this model, the soft threshold we impose on the eigenvalues can be incorporated in the eigenvalue distribution (Eq.~(\ref{model3})) by recognizing
\beqs
u(\lambda) = \frac{\lambda^2}{2c^2} + \frac{\xi}{1-\lambda}, 
\hspace{0.5cm}
g(\lambda) = \frac{\partial u}{\partial \lambda}  = \frac{\lambda}{c^2} + \frac{\xi}{(1-\lambda)^2}.
\eeqs 
Plugging in Tricomi's solution (Eq.~(\ref{eq:tricomi_sol})), and redefining the variable $x \equiv \lambda - a$, where $a$ is the left edge of the support, we obtain the spectral distribution,
\begin{widetext}
\beq \label{eq:after_tricomi_model3}
f(x) \equiv \rho(x+a) = \frac{1}{\pi \sqrt{x(l-x)}}
\left[
1 - \frac{1}{\pi}
\text{Pr}\int_0^l dx'
\frac{\sqrt{x'(l-x')}}{x - x'}
\left(
\frac{x' + a}{c^2} + \frac{\xi}{(1-a-x')^2}
\right)
\right]\,\mbox{.}
\eeq
\end{widetext}
where $l$ is the width of the support. 

The spectral distribution can be solved analytically. The first term of the Cauchy principle value in Eq.~(\ref{eq:after_tricomi_model3}) can be evaluated exactly, using 
\beqs
\text{Pr} \int_0^l dx' \frac{ \sqrt{x'(l-x')} }{x-x'}x' = 
-\frac{\pi}{8} \left( 
l^2 + 4 l x - 8 x^2
\right)
\eeqs
and
\beqs
\text{Pr} \int_0^l dx' \frac{ \sqrt{x'(l-x')} }{x-x'}a = 
a \pi (x - \frac{l}{2})\,\mbox{.}
\eeqs
The second term of the integral can only be evaluated when 
$l > 1 - a$, which always 
holds when $N$ is finite. The Cauchy principle value is
\beqs
\begin{split}
& \text{Pr} \int_0^l dx' \frac{ \sqrt{x'(l-x')} }{x-x'} 
\frac{\xi}{(1-a-x')^2} \\
& = 
-\frac{\pi\xi}{2}
\frac{(l-2x)(1-a)+lx}
{\sqrt{ (1-a)(1-a-l)} (1-a-x)^2}.
\end{split}
\eeqs
Denoting the gap between the right edge of the support and the critical eigenvalue $\lambda_\text{thres} = 1$ as $g_0$, and recognizing that $a = 1 - g_0 - l$, we obtain the spectral distribution (Eq.~(\ref{eq:after_tricomi_model3})) as given in the main text:
\begin{widetext}
\beqs
\rho_\text{soft}(\lambda) = \frac{1}{\pi\sqrt{(\lambda - 1 + g_0 + l)(1 - g_0 - \lambda) }}
\left[
1 + \frac{l^2}{8 c^2} + \left(
1 - g_0 - \frac{1}{2}l
\right) \frac{\lambda}{c^2}
-\frac{\lambda^2}{c^2}
+\frac{\xi}{2} 
\frac{ (2 g_0 - 2 g_0^2 + l - 2g_0 l)- (2g_0 + l)\lambda}{\sqrt{g_0(g_0+l)}(\lambda - 1)^2 }
\right]\mbox{.}
\eeqs
\end{widetext}
The only unknowns are the value of the gap $g_0$ and the width $l$ of the support, which we can solve for by setting $f(0) = f(l) = 0$, and obtaining
set of coupled equations
\beq \label{eq:coupled_al_new}
\begin{cases}
1 + \dfrac{l^2}{8c^2} + \dfrac{(1 - g_0 - l)l}{2c^2} + \dfrac{\xi l}{2} (g_0 + l)^{-3/2} g_0^{-1/2} = 0, \\
\\ 
1 - \dfrac{3l^2}{8c^2} - \dfrac{(1-g_0-l)l}{2c^2} - \dfrac{\xi l}{2} (g_0+l)^{-1/2} g_0^{-3/2} = 0.
\end{cases}
\eeq
This set of coupled equation gives us the scaling behavior of the gap $g_0$ and the correlation time $\tau_\text{corr}$ for different interaction strength $c$. Since the support $l$ shows up in many terms in the form of $l^2/c^2$, we can deduce that $l$ expands with the interaction strength $c$, as in both the GOE case and the GOE with hard threshold case. For plotting in the main text, the exact values of $g_0$ and $l$ are solved numerically.

\section{Compute autocorrelation function in Model 2}
The computation of autocorrelation function in Model 2 is subtle, since
\beqs
C_\text{hard} (t) = \int d\lambda  {{\rho(\lambda) }\over{(1-\lambda)}} e^{-(1-\lambda) |t|}\sim  \int dk\, k^{-1/2} \, k^{-1}\,e^{-k |t|}
\eeqs
is not integrable. After introducing an infrared (IR) cut-off at $ \varepsilon \sim g_0 \sim N^{-2}$, we can write the autocorrelation function as
\beqs
C_\text{hard}(t) \approx \int_{\varepsilon} dk k^{-3/2}e^{-k t} \\
=  -2 t^{1/2} \Gamma \left( \frac{1}{2}, \varepsilon t\right) 
+ 2 \varepsilon^{-1/2} e^{-\varepsilon t},
\eeqs
where $\Gamma(a,z)$ is the incomplete gamma function
\beqs
\Gamma\left( a , z \right) 
= \int_{z}^\infty u^{a-1} e^{-u} du.
\eeqs
The resulting autocorrelation coefficient is
\beq
\begin{split}
R_\text{hard}(t) = \frac{C_\text{hard}(t)}{C_\text{hard}(0)} & 
\approx e^{-\varepsilon t} - (\varepsilon t)^{1/2} \Gamma\left( \frac{1}{2}, \varepsilon t \right) \\
& = 1 - \sqrt{\pi} (\varepsilon t)^{1/2} + \varepsilon t  + \mathcal{O}((\varepsilon t)^2),
\end{split}
\eeq
and the correlation time 
\beq
\tau_\text{corr} = \frac{\int_\varepsilon dk \rho(k) k^{-2}}{\int_\varepsilon dk \rho(k) k^{-1}} \sim  \frac{\int_\varepsilon dk  k^{-5/2}}{\int_\varepsilon dk  k^{-3/2}} \sim\varepsilon^{-1} \sim N^2.
\eeq

\section{Scaling argument for finite gap in symmetric systems with soft threshold}

In this Appendix we will use scaling argument to show that in Model 3 (GOE with soft threshold constraining the norm activity), there is always a finite gap between the right edge of the support of the spectral density $\rho(\lambda)$ and the stability threshold $\lambda^* = 1$.

Let the interaction matrix $\bm M$ be symmetric. Recall Equation~(\ref{eq:soft_jointp_lambda}), the joint probability distribution for the eigenvalues is
\begin{widetext}
\beq
P(\lbrace\lambda_i \rbrace) \propto
\exp
\left[
-\frac{N}{2c^2}\sum_i\lambda_i^2
- N\xi \sum_i \frac{1}{(1-\lambda_i)^\beta}
+\frac{1}{2}\sum_{j\neq k}\ln \big\lvert \lambda_j - \lambda_k \big \rvert 
\right].
\eeq
\end{widetext}
Here, we use $\beta$ as the exponent for the purpose of generality (soft threshold has $\beta > 0$). For Model 3, $\beta = 1$. After coarse-graining and saddle point approximation, we obtain that the spectral density $\rho(\lambda)$ satisfies
\beq \label{eq:lambda_pr_general_beta}
\frac{\lambda}{c^2} + \frac{\beta \xi}{(1-\lambda)^{\beta+1}} = 
\text{Pr}\int \frac{d \lambda' \rho(\lambda')}{\lambda - \lambda'}.
\eeq
Let us denote 
$x \equiv 1 - \lambda$. If there is no gap between largest eigenvalue and $\lambda^* = 1$, we can write the spectral density near $\lambda^*$ as 
$\widetilde{\rho}(x) \equiv \rho(1 - x) =  x^\alpha$. Our maximum entropy constraint states that
\beqs
\begin{split}
\int d\lambda \rho(\lambda) \frac{1}{(1 - \lambda)^\beta} & = \text{const.} \\
\int_0 dx x^{\alpha - \beta} &  = \text{const.},
\end{split}
\eeqs
which implies $\alpha > \beta - 1$. 

For Equation~(\ref{eq:lambda_pr_general_beta}) to make sense, we need
\beqs
\frac{1 - x}{c^2} + \frac{\beta \xi}{x^{\beta + 1}} = \text{Pr}\int_0 \frac{d x' \widetilde{\rho}(x')}{x'- x} =  \text{Pr}\int_0 \frac{d x' x'^\alpha}{x'- x}  = x^\alpha.
\eeqs
As $x\rightarrow 0$, this equation requires $\beta + 1 = -\alpha$, for both sides to have the same speed of divergence.  This requires $\beta < 0$, which is contradictory to the soft threshold assumption. 

This contradiction means there has to be a non-vanishing gap $g_0$ between the largest eigenvalue and the stability threshold $\lambda^*$. We can re-write the density function using 
$y = 1 - g_0 - \lambda$, then Eq.~(\ref{eq:lambda_pr_general_beta}) becomes
\beq
\frac{1 - g_0 - y}{c^2} + \frac{\beta \xi}{(y + g_0)^{\beta + 1}}
 = \text{Pr} \int_{0}
 \frac{dy' y'^\alpha}{y' - y}.
\eeq
The right hand side does not explicitly depend on $g_0$, so we have
\beq
g_0 \sim \xi^{1/(\beta + 2)}.
\eeq
For Model 3, since $\beta = 1$, the gap $g_0 \sim \xi^{1/3}$.

\section{Derivation for the scaling of time constants in Model 3}
\label{sec:lts_appd:scaling_calc}

For the ensemble of connection matrices with a maximum entropy constraint on the norm activity (Model 3), we are interested in how the time constants, $\tau_\text{max}$ and $\tau_\text{corr}$, depend on the parameters of the system: the interaction strength $c$, and the constrained norm activity $\mu$. In particular, how the time constants scale with the norm activity when the system is set in different phases by the interaction strength $c$. Here, we analyze the spectral distribution $\rho(\lambda)$, and investigate how the gap size $g_0$, the length of the support $l$, and the constrained norm activity $\mu$ scales with the Lagrange multiplier that we used to constrain the norm, $\xi$. The results are summarized in Table~\ref{tab:scaling}, and can be visualized in Fig.~\ref{fig:soft_mf}.

We have shown that the spectral distribution is (also see Eq.~(\ref{eq:model3_spectrum}))
\begin{widetext}
\beq
\begin{split}
\rho(\lambda) & = \frac{1}{\pi\sqrt{(\lambda - 1 + g_0 + l)(1 - g_0 - \lambda) }} B(\lambda)
\mbox{,} \\
B(\lambda) & = \left[
1 + \frac{l^2}{8 c^2} + \left(
1 - g_0 - \frac{1}{2}l
\right) \frac{\lambda}{c^2}
-\frac{\lambda^2}{c^2}
+\frac{\xi}{2} 
\frac{ (2 g_0 - 2 g_0^2 + l - 2g_0 l)- (2g_0 + l)\lambda}{\sqrt{g_0(g_0+l)}(\lambda - 1)^2 }
\right].
\end{split}
\eeq
\end{widetext}
By setting the spectral density at the edge of the support to zero, we can solve for the scaling dependence of gap size $g_0$ and the length of the support $l$ on the Lagrange multiplier $\xi$. Mathematically, we have
\beq
\begin{split}
B(1 - g_0 - l) & = 0, \\
B(1 - g_0) & = 0.  
\end{split}
\eeq
After simple algebraic manipulation, this set of constraints becomes
\beqn \label{eq:soft_edge1}
(2g_0 + l)(8 c^2- 3l^2) + 4 l^2 = 0,\\
\label{eq:soft_edge2}
8 - \frac{l^2}{c^2} - 2\xi l^2 (l + g_0)^{-3/2} g_0^{-3/2} = 0, 
\eeqn
which sets the scaling relation between the longest time scale $\tau_\text{max} = 1/g_0$ and the Lagrange multiplier $\xi$.

For the correlation time $\tau_\text{corr}$, we can express it as the ratio
\beq
\tau_\text{corr} = \frac{\nu}{\mu},
\eeq
where the denominator is the expectation value for averaged norm activity,
\begin{widetext}
\beq \label{eq:norm_soft}
\begin{split}
\mu = \langle x_i^2\rangle & = \int d\lambda \rho(\lambda) \frac{1}{1-\lambda} \\
& = \frac{1}{c^2} + \frac{1}{c^2 \sqrt{g(g+l)}}\left( c^2 - g - \frac{l}{2} + \frac{l^2}{8} \right) -\frac{\xi}{8}\frac{l^2}{g^2(g+l)^2},
\end{split}
\eeq
and the numerator can be written as 
\beq
\begin{split}
\nu & \equiv \int d\lambda \rho(\lambda) \frac{1}{(1-\lambda)^2} \\
& = -\frac{1}{c^2} + \frac{1}{c^2}g^{-3/2}(g+l)^{-3/2} (c^2 g + g^3 + \frac{1}{2} c^2 l + \frac{3}{2}g^2 l - \frac{1}{4}l^2 + \frac{5}{8}g l^2 + \frac{1}{16}l^3) -\frac{\xi}{8}\frac{l^2(2g+l)}{g^3 (g+l)^3}.
\end{split}
\eeq
\end{widetext}

\begin{table*}[htb]
\centering
\begin{tabular}{| c | c | c | c | c |}
\hline
 & & $g_0$ & $l$ & $\langle x_i^2\rangle$ \\ \hline
 $\xi \ll 1 $ & $ 0< c < 1/\sqrt{2} $ & $1 - \sqrt{2}c  + A_{-}\xi$ & $2\sqrt{2}c - B_{-} \xi$ & $\frac{1}{c^2} - \frac{\sqrt{1 - 2c^2}}{c^2}- D_{-}\xi$ \\ \hline 
 $\xi \ll 1 $ & $c = 1/\sqrt{2}  $ & $A_c \xi^{2/5}$ & $2 - B_c \xi^{2/5}$ & $2 - D_c \xi^{1/5}$ \\ \hline 
$\xi \ll 1 $ & $c > \sqrt{2} $ & $A_{+} \xi^{2/3}$ & $l_0 - B_{+} \xi^{2/3}$ & $D_{+} \xi^{-1/3} + \frac{1}{c^2}$ \\ \hline 
%$\xi \gg 1 $ & $c>0 $ & $\sim \xi^{1/3}$ &  & $\sim %\xi^{-1/3}$ \\
%\hline
\end{tabular}
\caption{Scaling of inverse slowest time scale (gap) $g_0$, width of the support of spectral density $l$, and averaged norm per neuron $\langle x_i^2\rangle$ versus the Lagrange multiplier $\xi$ (to leading order) in different regimes. }\label{tab:scaling}.
\end{table*}

Because in the limit of $\xi \gg 1$, $g_0$ is large and there is no long time scales, we focus our discussions on cases where $\xi \ll 1$.
 
\textbf{Case 1: $c< 1/\sqrt{2}$} - 
As $\xi$ approaches 0, we recover the semicircle spectral density with the wall far away from the spectrum. This suggests we can write $g_0 = 1 - \sqrt{2}c + A_{-}\xi^{\gamma} $ and $l = 2\sqrt{2} c - B_{-} \xi^\gamma$. The expected value for the norm is
\beq
\begin{split}
\lim_{\xi\rightarrow 0} \langle x_i^2 \rangle\vert_{c < 1/\sqrt{2}} & =
\int_{-\sqrt{2}c}^{\sqrt{2}c} d\lambda \frac{1}{c\pi}\frac{\sqrt{2-\lambda^2/c^2}}{1 - \lambda} \\
 & = \frac{1}{c^2} - \frac{\sqrt{1 -2c^2}}{c^2}.
\end{split}
\eeq
Taylor expansion leads to $\gamma = 1$. As $\xi$ decreases, both $g_0$ and $\mu$ reaches an upper limit. There is no long time scales in this case.

\textbf{Case 2 $c = 1/\sqrt{2}$} - This is the critical case in the hard wall limit, when the gap $g_0$ is 0, and the spectral density remains a semicircle. To solve for the scaling behavior when $\xi \ll 1$, we follow similar procedure as in the previous section and assume $g_0 = A_{c} \xi^\gamma$, $l = l_0 + B_c\xi^\gamma$. And again, $l_0 = 2\sqrt{2}c = 2$. Now, the 0th order terms of Eq.~(\ref{eq:soft_edge2}) has no terms that scales with $\xi$, requiring us to go to higher orders to solve for $\gamma$,
which gives
\beq
\gamma = \frac{2}{5}.
\eeq
The norm is 
\beq
\begin{split}
\langle x_i^2 \rangle = 2 - D \xi^{1/5} + \mathcal{O}(\xi^{3/5}).
\end{split}
\eeq
This is interesting: as $\xi$ decreases, the norm activity $\mu$ approaches the limit $\mu = 2$, and the corresponding $g_0$ continuous to decrease. The system can reach an infinitely long timescale with a bounded dynamic range for individual neurons.

\textbf{Case 3 $c > 1/\sqrt{2}$} - 
In this regime, the spectral density at $\xi = 0$ is divergent near $\lambda = 1$. But for any $\xi > 0$, there is a finite gap between the wall and right edge of the spectrum, so a limit for the spectral density when $\xi \rightarrow 0$ is not well-defined.

We can assume the gap $g_0$ takes the form $g_0 = A_{+} \xi^\gamma$. Because $l$ is of order 1, we can assume $l = l_0 + B_{+} \xi^\gamma$.  Plugging these to Eq.~(\ref{eq:soft_edge1}), we can solve the 0th order equation and get 
\beq
l_0 = \frac{2}{3}\left(
1 + \sqrt{1 + 6c^2}
\right).
\eeq
This $l_0$ is equal to the length of the spectrum at the hard wall limit, i.e. when $\xi = 0$. Plugging $g_0$ and $l$ into Eq.~(\ref{eq:soft_edge2}), we get from 0-th order solution
\beq
\mu = \frac{2}{3}
\eeq
and
\beq
A_{+} = 2^{-2/3}l_0^{1/3}(2-\frac{l_0^2}{4c^2})^{-2/3}.
\eeq
We notice that the prefactor $A$ is not a monotonic function of $c$. Rather, $A$ takes a minimum $A_\text{min} = 1$ at $c = 2$. This is interesting, as we don't want the interaction strength to be too large.

The norm becomes
\beq
\mu = \int d\lambda \rho(\lambda) \frac{1}{1-\lambda} = D_{+} \xi^{-1/3} + \frac{1}{c^2} + \mathcal{O}(\xi)^{1/3},
\eeq
where 
\beq
D_{+} = \frac{1}{c^2} A^{-1/2} l_0^{-1/2} \left( c^2 + \frac{l^2}{8} - \frac{l}{2} \right) - \frac{A^{-2}}{8}.
\eeq
Similarly, the prefactor $D_{+}$ has a maximum $D_\text{max} = 3/8$ at $c = 2$. 

For the correlation time, the denominator is the averaged norm, and the numerator is 
\beq
\begin{split}
\nu & \equiv \int d\lambda \rho(\lambda) \frac{1}{(1-\lambda)^2} \\
& = E\xi^{-1} -\frac{1}{c^2} + \mathcal{O}(\xi).
\end{split}
\eeq
The prefactor $E$ has maximum $E_\text{max} = 1/8$ at $c = 2$. Then, the correlation time is
\beq
\tau^\text{sw}_\text{corr} = \frac{E}{D}\xi^{-2/3} + \mathcal{O}(\xi).
\eeq
Interestingly, when we compute the ratio of the two time scales, $\tau_\text{max}$ and $\tau_\text{corr}$, we found that the ratio is a constant, as 
\beq
\frac{\tau_\text{corr}^\text{sw}}{\tau_\text{max}^\text{sw}} = \frac{1}{3} - \frac{A E}{c^2 D^2} \xi^{1/3} + \mathcal{O}(\xi).
\eeq
In this case, as $\xi$ approaches 0, the gap $g_0$ decreases, but the norm also increases to infinite. If we only look at the relation between the parametrized variable $g_0$ and $\mu$, we find that $g_0 \sim \mu^{-2}$, which matches what we get from numerically solving the exact Eqs.~(\ref{eq:soft_edge1}) and~(\ref{eq:soft_edge2}). To get an infinitely long time scale, an unlimited growth of dynamic range for individual neuron is required, which is impossible to realize in biological systems.

\section{Asymmetric interaction}

Recall that the system we consider is a linear dynamical system
\beq
\frac{d \dot{x}_i}{dt} = -x_i + \sum_j M_{ij} x_j + \eta_i(t),
\eeq
with the entries of the interaction matrix $\bm{M}$ real, and white noise $\bm{\eta}$ where
\beqs
\langle \eta_i(t)\rangle = 0,
\hspace{0.5cm}
\langle \eta_i(t) \eta_j(t') \rangle = 2\delta_{ij}\delta(t-t').
\eeqs
If $\bm M$ can be asymmetric, the solution of this dynamical system gives the steady state autocorrelation at equal time
\beq
\mu = \frac{1}{N}\big \langle \sum_i x_i^2 \big \rangle
 = \frac{1}{N}\sum_{i,j}O_{ij}\frac{2}{2 - \lambda_i - \lambda_j^\dagger}.
\eeq
Here, $O_{ij}$ is the overlap between the left and right eigenvectors, defined as
\beq
O_{ij} = \langle l_i \vert l_j \rangle
\langle r_j \vert r_i \rangle,
\eeq
where $\langle l_i \vert$ is the left eigenvector (and $\vert r_i \rangle$ is the right eigenvector) corresponding to the $i$-th eigenvalue. The normalization is chosen such that the left and right eigenvectors form a bi-orthogonal basis, i.e. $\langle l_i \vert r_j \rangle = \delta_{ij}$. 

In Model 3, we consider systems which is stable by constraining the autocorrelation when the system is in steady state. The corresponding maximum entropy distribution for the matrix element is
\beq \label{eq:asym_measure_matrix}
P(\bm M) \propto \exp\left[-\frac{N}{2c^2}\text{Tr} \bm M^\intercal \bm M 
- N\xi \sum_{i,j} O_{ij} 
\frac{2}{2-\lambda_i-\lambda_j^\dagger}
\right],
\eeq

\begin{figure}[ht]
\includegraphics[width=0.99\columnwidth]{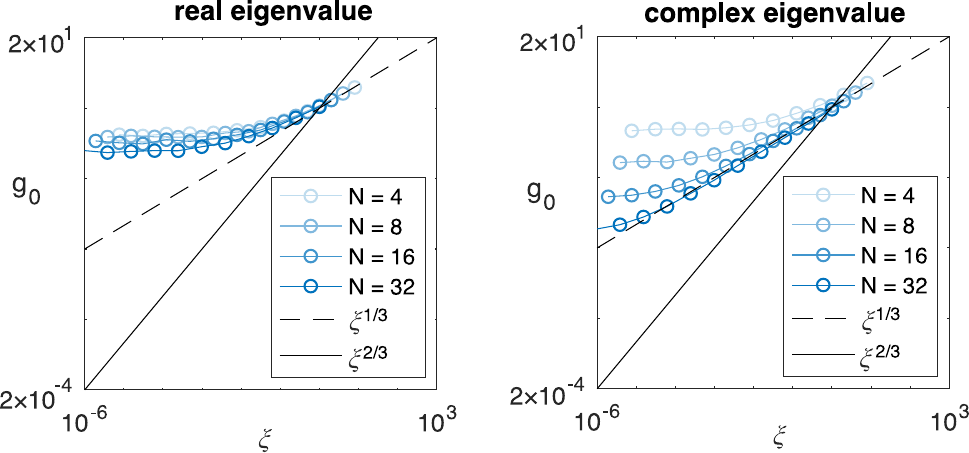}
\caption{Relation between the gap $g_0$ and the Lagrange multiplier $\xi$ for real asymmetric matrix $\bm{M}$ subject to constraints on the norm. The connection strength $c = 2$. Slow modes are in general more likely to be found in eigenvalues which form complex conjugate pairs (\textit{right}), than real eigenvalues (\textit{left}), suggesting slow modes are more likely oscillatory than exponential decay.}\label{fig:real_asym_g0_vs_t}
\end{figure}

This probability measure of the matrix can no longer be separate into the product measure of eigenvalue and eigenvectors due to the term with eigenvector overlaps. Another source of complexity comes from additional constraints for eigenvalues of real asymmetric matrices, where all eigenvalues have to be real or form complex conjugate pairs. 

To get an idea about whether the gap in the symmetric case remains in asymmetric systems, we directly sample matrices from the distribution [Eq. (E4)]
using Monte Carlo Markov Chains. As shown in Fig.~6,
there is still a finite gap $g_0$ between the right edge of the spectrum and the stability threshold, the same as in the symmetric case. It is interesting to notice that the real eigenvalues and the complex eigenvalues behave differently: the gap between the complex eigenvalue and the wall can be fitted to a $g_0 \sim \xi^{1/3}$ curve, while the gap between the largest real eigenvalue and the wall seems to be independent of $\xi$, suggesting that slower modes are more likely to be found oscillatory than exponential decay. Other interesting thing to notice is that this $1/3$ scaling is slower than what we got from symmetric systems.

\end{document}